\documentclass[prb,preprint]{revtex4-1} 

\usepackage{amsmath}  
\usepackage{amsfonts} 
\usepackage{graphicx} 
\usepackage{braket} 
\usepackage{enumitem}
\usepackage{upgreek}
\usepackage{pgfplots}
\begin{document}

\title{Quantum Key Distribution using Expectation Values \\of Super-classical GHZ States}

\author{Hyung S. Choi}
\email{hyung.choi@greenville.edu} 
\author{Ye Jin Han}
\author{Collin Kessinger}
\author{Qiaoren Wang}
\affiliation{Department of Physics, Greenville University, 315 E College Ave, Greenville, IL 62246}


\begin{abstract}
We propose a new quantum key distribution scheme that is based on the optimum expectation values of maximally entangled Greenberger-Horne-Zeilinger states. Our protocol makes use of the degrees of freedom in continuously variable angles, thereby increasing the security of the key distribution. Outlined are two protocols that distribute a key from Alice to Bob using the above idea, followed by an extension that allows for the same key to be shared with Charlie. We show how this scheme provides for certain detection of any eavesdropper through absolute violation rather than the probabilistic violation used in many protocols.
\end{abstract}
\maketitle 

\section{Introduction} 
\subsection{Cryptography and Quantum Key Distribution}
Current cryptographic schemes have been threatened by the looming development of quantum computers and potential mathematical breakthroughs in the understanding of P vs. NP problems. This fundamental realization of computation has already proposed a method to break RSA encryption using Shor's algorithm.\cite{shor} With RSA being the most commonly used public encryption algorithm, creating a new quantum secure encryption is essential. Currently, the only protocol that is completely quantum and classically secure is the Vernam cipher, more commonly known as the one-time pad. To use the one-time pad, a key consisting of random bits of equal length to the message must be distributed. This key is then XORed with the message and the result is sent classically to a recipient who possesses the distributed key for decoding. The used key is then discarded and any further communication requires the distribution of another distinct, random key. In order to achieve secure communication, secure key distribution is critical. Quantum Physics enables sending a key without the use of another key via quantum key distribution (QKD). In this paper, we propose a new QKD protocol that enables certain detection of any potential eavesdropper. We make use of three particle entangled Greenberger-Horne-Zeilinger (GHZ) states that can safely send a key and quickly detect an eavesdropper by making use of optimum spin or polarization expectation values. Instead of making use of  Bell's probabilistic inequality, we make use of a complete violation between the classical and quantum mechanical case.

\subsection{BB84 and Ekert91 Protocols}
The BB84 protocol is a widely known QKD method that uses a string of superposed unentangled qubits. In this scheme, Alice measures 4n qubits in either the X or Z basis randomly. Alice then sends these qubits to Bob, who performs his own measurements, ignorant of the basis chosen by Alice for any particular qubit. Bob and Alice can then announce the basis they chose for each qubit and discard qubits that were measured differently (2n). Alice then announces the measurement results of half of the remaining bits (n). Bob and Alice are now able to compare and determine if an eavesdropper was present. In the end, they share a completely secret key of length n if no eavesdropper was detected.

It has been shown that maximally entangled Bell States can be used to show violation between classical and quantum physics. This violation is realized through probability and requires a sufficient number of pairs to be tested in order to acquire accurate probabilities. The Ekert91\cite{ekert} protocol makes clever use of this probabilistic violation for QKD. Both Alice and Bob are in possession of one qubit of n maximally entangled bell state systems. Alice then takes secret measurements on all qubits using a basis chosen randomly between $0, \pi/8, \pi/4.$ Bob does the same but between the bases $0, \pi/8, -\pi/8.$ They then announce which basis they used for each qubit. They keep the qubits that were measured using the same basis. The remaining qubits are used to test for Eve. If Eve was present, she would have introduced local realism, indicated by significant deviation from the theoretical quantum probability coefficients calculated between the different bases used by Alice and Bob for a particular pair. If no eavesdropper was present, Alice and Bob now share a secret key of length n. Note that this scheme provides for probabilistic detection of Eve. 

\subsection{Quantum Key Distribution Schemes Using Maximally Entangled GHZ States}
Mutipartite cryptographic schemes using GHZ and W-states are not novel ideas, and methods to superdense code with GHZ states have already been proposed.\cite{anne}$^{,}$\cite{hao} Cao Hai-Jing and Song He-Shan\cite{cao} present a theoretical QKD scheme with four particles based on entanglement swapping. Their protocol users share keys by conducting bell state measurements and sending results via a classical channel. Liu et. al\cite{liu} demonstrates a multipartite quantum secret sharing protocol based on a symmetric W-State and requires only a single photon measurement in order to recover the key from the sender. Joo et. al\cite{joo} proposes a three particle entangled W-state protocol that makes use of local measurements on particles with different axes. 

Xing-Ri Jin\cite{jin} et. al presents a protocol for simultaneous  quantum secure direct communication (QSDC) amongst $n$ members using $n$ particles in GHZ states. The message that the party members want to send out simultaneously is encoded via unitary operations, and an eavesdropper is detected by measuring and checking the error rate.

T. Gao presents two different QSDC schemes with GHZ states. In his first protocol, Gao et. al\cite{gaoal} presents a deterministic QSDC scheme that uses entanglement swapping and a three particle GHZ state as a quantum information channel. Local and predetermined unitary operations are performed on the particles in order to encode the message. In Gao's\cite{gao} second protocol, the sender directly encodes their message on a sequence of particles in the GHZ state and transmits it via quantum teleportation. In both of Gao's protocols, the eavesdropper is detected by testing the security of the quantum channel.

Kai Chen \& Hoi-Kwang Lo\cite{kai} presents an exhaustive description and study of a multi-partite QKD protocol using GHZ and W-States. Their protocol is unique in that even if the initial states fail to violate the standard Bell inequalities, their cryptographic scheme still remains secure. Their multipartite key agreement involves a two-way classical communication that achieves a secure key distribution for W-states when specific parameters are met. Whenever the members of the protocol share a generalized W-state, they achieve a secure conference key agreement. Their scheme functions not only with GHZ states, but also with CSS states. In the case that Eve imposes as one of the senders, Kai and Lo provide a depolarizing procedure for the ancilla bits held by the eavesdropper. Other quantum cryptographic schemes involving GHZ states are cited in the bibliography\cite{faisal}$^{,}$\cite{zeng}$^{,}$\cite{hillery}$^{,}$\cite{tittel}$^{,}$\cite{saber}$^{,}$\cite{wang}.

Contrary to the the mentioned QKD schemes using GHZ or W-states, our proposed protocol makes use of GHZ states and optimum expectation values of spins or retarders at variable angles and does not involve unitary operations on the entangled particles. 
The addition of an angle makes it impossible for Eve to determine the correct measurement, and the moment Eve tampers with the entangled particles, she will be certainly detected. When the angles are set within certain parameters, the eavesdropper can be detected via an apparent and clear violation between a classical and quantum mechanical measurement instead of Bell's probabilistic inequality. A similar QKD protocol was presented by Mermin\cite{mermin} who used specific angles to produce similar results.

\section{Super-classical GHZ States} 
It is generally impossible to classically simulate quantum physics. In very specific cases, one can achieve a super-classical generalized EPR state, meaning the expectation value is always equal to $\pm1$, and it is impossible to duplicate this result classically. In our scheme, we further expand on Greenberger, Horne and Zeilinger's\cite{ghz} calculation of expected spin values that lead towards a super-classical case when certain parameters are met.

\subsection{Expectation value for Spins}

Satisfying the condition 
$$\Upsilon(\theta) = +\ket{\theta}\bra{\theta} - \ket{\theta_\bot}\bra{\theta_\bot} $$

\noindent we derive the spin-measurement operator along any polar and azimuthal angle where $\hat{n}$ is the direction of the measurement of the spin as 

\begin{equation}
\Upsilon \equiv \vec{\sigma}\cdot \hat{n} = \begin{pmatrix}
 															    \cos{\theta} & \sin{\theta}\exp(-i \phi)  \\
    															    \sin{\theta}\exp(i \phi) & -\cos{\theta}
														      	\end{pmatrix}
\end{equation}

\vspace{5mm}
The expected spin value of a particle can be found by using the equation

$$ E(n) = \braket{\psi|\vec{\sigma}\cdot \hat{n}|\psi} $$

The expectation value of a spin measurement of more than one particle in different directions is given by 

\begin{equation*}
E(n_{1}, n_{2},\dots,n_{x}) = \braket{\psi| \Upsilon_{1} \otimes \Upsilon_{2} \cdots \Upsilon_{x}|\psi}
\end{equation*}

\noindent where $x$ is the number of particles and $n$ is the direction in which that particle is measured. We find the expectation value of a three particle GHZ state $\ket{\psi}  = \frac{1}{\sqrt{2}} \left( \ket{+++} - \ket{---} \right) $  when $\theta = \frac{\pi}{2}$ is

$$ E = -\cos(\phi_{1} + \phi_{2} + \phi_{3})$$

In general, the expectation value of a three particle GHZ state when $\theta = \frac{\pi}{2}$ is given by

\begin{equation}
E = \pm \cos(\pm\phi_{1} \pm \phi_{2} \pm \phi_{3})
\end{equation}

If the superposition is added, the cosine is positive and viceversa. For example, $\frac{1}{\sqrt{2}} \left( \ket{+++} - \ket{---} \right)$ yields an expectation value of $-\cos(\phi_{1} + \phi_{2} + \phi_{3})$ while  $\frac{1}{\sqrt{2}} \left( \ket{+++} + \ket{---} \right)$ yields $+\cos(\phi_{1} + \phi_{2} + \phi_{3})$. Similarly, the signs of the left hand side of the entangled particles determine the sign of the $\phi$'s. $\frac{1}{\sqrt{2}} \left( \ket{++-} - \ket{--+} \right)$ yields an expected value of $-\cos(\phi_{1} + \phi_{2} - \phi_{3})$ while $\frac{1}{\sqrt{2}} \left( \ket{-++} - \ket{+--} \right)$ yields  $-\cos(-\phi_{1} + \phi_{2} + \phi_{3})$. This way one can quickly determine the expected value for any three particle GHZ state. 

When $\pm\phi_{1} \pm \phi_{2} \pm \phi_{3} = 0, \pi$ and the spin measurement expectation value yields $+1$ we can determine with certainty that the measurement will collapse to $\ket{+++}, \ket{+--}, \ket{-+-},$ or $ \ket{--+}$ with equal probability, as the product of the spins is equal to $+1$. Consequently, the same can be said when the expectation value is $-1$. In such case, the wave function will collapse to $\ket{---}, \ket{++-}, \ket{+-+},$ or $ \ket{-++}$. Within these parameters, Greenberger, Horne and Zeilinger\cite{ghz} noted that if one knows the spin states of two particles and the expectation value is $\pm1$, the third can be predicted with 100\% certainty.

\subsection{Expectation Value of Polarization}
The same result can be achieved when measuring the expected polarization value using retarders. A retarder with its azimuthal angle $\rho = 0$ and retardance $\delta$ can be expressed as
\begin{equation}
W = \begin{pmatrix}
			\exp(-i\frac{\delta}{2}) & 0 \\ 
			0 & \exp(i\frac{\delta}{2})
\end{pmatrix}
\end{equation}

Satisfying the condition 
$$T(\theta) = +\ket{\theta}\bra{\theta} - \ket{\theta_\bot}\bra{\theta_\bot} $$

\noindent the polarization measurement operator along angle $\theta$ is given by: 
\begin{equation}
T(\theta) = \begin{pmatrix} 
             \cos{2\theta} & \sin{2\theta}\\
             \sin{2\theta} & -\cos{2\theta}
\end{pmatrix}
\end{equation}

The polarization measurement of a particle with retarders can then be found using
\begin{equation}
\Lambda = W^\dagger T W = \begin{pmatrix}
							\cos{2\theta} & \sin{2\theta}\exp(i\delta)\\
								 \sin{2\theta}\exp(-i\delta) & -\cos{2\theta}
\end{pmatrix}
\end{equation}

We then calculate the expectation value of the coincident measurement of entangled particles  by computing 
$$E= \braket{\psi| \Lambda_{1} \otimes \Lambda_{2} \cdots \Lambda_{i}|\psi}$$
\noindent where $i$ is the number of particles. For three entangled particles in a general GHZ state with $\theta = \frac{\pi}{4}$, we get

\begin{equation}
E = \pm \cos(\pm\delta_{1} \pm \delta{2} \pm \delta{3})
\end{equation}

Just like the spin case, whenever $\pm\delta_{1} \pm \delta{2} \pm \delta{3} = 0, \pi$, we get a super-classical GHZ state with which one can deduce the third particle's polarization if the polarization measurements of the other two particles are given. The plus and minus signs follow the same pattern as with the spin case. If we have $\frac{1}{\sqrt{2}} \left( \ket{+-+} - \ket{-+-} \right)$, the expectation value will be  $-\cos(\delta_{1} - \delta_{2} + \delta_{3})$.

\section{Quantum Key Distribution Scheme}
Using super-classical GHZ states, we now present two QKD methods. Here, a measurement of $\ket{+}$ corresponds to the 0 bit and $\ket{-}$ corresponds to the 1 bit. In our protocol, three parties (Alice, Bob, and Charlie) work together to securely send a randomly generated key $K$ from Alice to Bob. We then briefly explain an extension of either protocol that allows the same key to also be distributed to Charlie. We denote the XOR operation with $\oplus$ and XNOR operation with $\odot$.

\subsection{Method 1}
\begin{enumerate}[noitemsep,nolistsep]
  \item Bob prepares entangled particles $a, b, c$ in a specific GHZ state and determines the formula for calculating the expectation value. He then declares through a public channel three different angles (retardance values) from which Alice, Charlie, and Bob can choose from to measure their particle's spin or polarization. It is ideal to select three unique angles (retardance values) that minimize the number of combinations that do not sum to either 0 or $\pi$.
  \item Bob sends particle $a$ to Alice and $c$ to Charlie. The three participants then take the joint measurement of their particles using one of three allowed values randomly. The resulting measurements are labeled A,B,C accordingly.
  \item Alice and Charlie announce which angles (or retardance values) they chose through the public channel.
  \item Bob computes $\pm \cos(\pm\phi_{a} \pm \phi_{b} \pm \phi_{c})$ (using the form that relates to the GHZ state that he prepared). If the expectation value does not equal $\pm 1$, Bob tells Alice to discard that qubit through a public channel. 
  \item Alice takes $A\oplus K\equiv D$.
  \item Both D and C are announced through the public channel.
  \item Bob takes $D\oplus C\equiv E$.
  \item Bob takes $E\oplus B\enspace (E\odot B \enspace for E= -1)$. The result is the key that Alice generated.
  \item Bob is then able to determine A by taking $K\oplus D$. Bob uses only qubits from measurements with an expectation value of either $\pm 1$, not both, for detection of Eve. If the total measurement is not equal to +1 (-1 for E = -1) then Bob immediately knows that Eve was present.
\end{enumerate}

\subsection{Method 2}
\begin{enumerate}[noitemsep,nolistsep]
\item Bob generates any GHZ state and sends qubits $a$ and $c$ to Alice and Charlie respectively. The GHZ state can be generated by any party, however, it may be ideal for Bob to generate the state. We will see later that Bob is the only party that needs to know which of the eight unique GHZ states has been generated.
\item Alice and Charlie then choose and announce publicly which angle (or retardance value), $\phi_a$ and $\phi_c$ respectively, they will use to measure.
\item Bob then sets $\phi_b$ in such a way as to produce either $\pm 1$ to his preference. Throughout the process Bob will ensure that the expectation value is always the same as the original choice.
\item All three parties then take the joint-measurement according to their respective angles (retardance values). The resulting measurements are labeled A,B,C accordingly.
\item Alice then takes $A\oplus K\equiv D$
\item D and C are announced publicly.  
\item Bob then takes $D\oplus C\equiv E$
\item Bob then takes $E\oplus B\enspace(E\odot B\enspace$ for $E=-1)$. The result is the original key that Alice generated.
\item Bob is then able to determine A by taking $K\oplus D$. This will allow him to check for the presence of an eavesdropper. If the total measurement is not equal to 1 (-1 for E =-1) then Bob immediately knows that Eve was present.

\end{enumerate}
\subsection{Example of Protocol}
Below is a chart containing an example of either protocol. In this example, four bits are being sent from Alice to Bob. The chart is not laid out in order of information generated, but rather in order of information used in the protocol. The measurement values were chosen arbitrarily and do not come from any real experiment. They do, however, follow the case of an expectation value of +1 and the absence of an eavesdropper. 

Note that an expectation value of +1 indicates that between A, C, and B there \textit{must} be either a zero or an even number of ``1"s. The result (K) is possessed by Bob and is the original key which Alice wishes to send.
\newline

\begin{tabular}{c c c c c c c}
$KEY\equiv K$ (known by Alice)&1&0&1&1\\
A (known by Alice and determinable by Bob at the end)&0&1&1&0\\
$A\oplus K\equiv D$ (public)&1&1&0&1\\
C (public)&0&0&1&0\\
$D\oplus C\equiv E$ (public, calculated by Bob)&1&1&1&1\\
B (known by Bob)&0&1&0&0\\
$E\oplus B=K$ (calculated by Bob, used to determine A for Eve detection)&1&0&1&1

\end{tabular}
\vspace{5mm}

\subsection{Extension to Include Key Distribution to Charlie}
In both of the methods outlined above, Charlie does not receive the key, and Alice and Bob do not need to trust Charlie. This will be shown in the following section entitled Detection of Eve. We now briefly outline an extension of either of the above protocols that allows the same key to be additionally distributed to Charlie.

All three parties first run either of the two protocols which results in a shared random key held by both Alice and Bob. The three parties then switch roles, with Charlie acting as Bob, Bob acting as Alice, and Alice acting as Charlie. Either protocol is implemented and Bob (acting as Alice) sends the shared key to Charlie (acting as Bob). Now all three parties share the same random key. 

This does not decrease the security of the protocol in any way. The only public information shared between the two protocols that could jeopardize the secrecy of the key is the key XORed with Alice's (Bob's for the second distribution) measurement seen below.

$$A_1\oplus K=D_1$$ 
$$A_2\oplus K=D_2$$

Both $D_1$ and $D_2$ are known publicly while $A_1$, $A_2$, and $K$ are secret. Initially, this seems to be in violation of the requirements of the one-time pad. However, since $A_1$, $A_2$, and $K$ are quantum random, this extension does not violate the requirements of the one-time pad. Thus, the intersection between multiple runs of the protocol is secure.

\section{Detection of Eve}
Bob is in possession of the expectation value of the system and all three measurement results. Since he set the expectation value to either $\pm 1$, the total result will be that value. If there is no noise in the system, Bob knows that the total value will not equal the expectation value only if Eve was present. If Eve is present, she will cause absolute violation 50\% of the time. Realistically, there will be noise in all practical systems, so it is necessary to determine a noise threshold for any system and take the average total measurement over several runs. If this threshold is violated, Bob knows Eve was present. We have shown that our protocol provides for absolute detection of Eve when accounting for noise in the system.

Let us explore the case in which Eve intercepts the particle intended for Charlie and carries out all actions normally required of Charlie, which includes setting and announcing $\phi_c$ and announcing the measurement of particle c: in this case, Eve has access to the information $\phi_a,\phi_c$, D, and C. This is not enough information to determine the message. To do so, Eve would need to additionally obtain either A or B. While A is encoded in D, the information contained is as secure as the one time pad because both K and A are random. The knowledge of $\phi_a, \phi_c$, and C is not enough to determine the effect Alice's measurement had on the key bit. Eve would know that the key bit was flipped with a probability $\frac{1}{2}$. Therefore, in the event of Eve imposing as Charlie, our scheme is still as secure as the one-time pad. It is worth noting that it is not necessary for Alice and Bob to trust Charlie. 

We now discuss security details uniquely pertaining to Method 1. In the particular case where an eavesdropper taps into Method 1's protocol and tries to measure particle $a$ before it reaches Alice, Eve will have to guess between the three angles for measurement that Bob announced. If Eve were to choose the same angle as Alice, the entangled particles will immediately collapse, so even if Alice still measures with the same angle, the observed value of the three particles will violate the expectation value 50\% of the time. After Bob has received and decoded the bits that Alice has sent, Bob can use the bits that yielded an expectation value of either $+1\enspace or -1$ (but never both) to see how often the measurement and prediction yield the same result. If Bob concludes that Eve is present, he can decide to start the whole process over again. If Bob concludes that Eve has not been tapping in, Alice, Bob and Charlie will continue the protocol.

\section{Concluding Remarks}
The quantum cryptographic scheme we propose is based on the expectation values of super-classical GHZ states. Contrary to the current quantum cryptography and key distribution using probabilistic comparison, our methods make use of the expectation value of spin or polarization, giving us an extra layer of security and making it impossible for Eve to retrieve the message. By the violation between classical and quantum case, Bob can immediately know if there exists an eavesdropper. This method can also expand to a three-party key distribution without decreasing the security of the scheme in any way. We present our encryption and decryption algorithm under the assumption that our particles are sent via a theoretically ideal, noiseless channel. We hypothesize that experimental realizations of this scheme accounting for noise will lead to complete detection of Eve.


\end{document}